# Infrared fingerprints of few-layer black phosphorus

*Guowei Zhang[1,2], Shenyang Huang[1,2], Andrey Chaves[3,4], Chaoyu Song[1,2], V. Ongun Özçelik[5], Tony Low[6] & Hugen Yan[1,2]\**

[1]Department of Physics, State Key Laboratory of Surface Physics and Key Laboratory of Micro and Nano Photonic Structures (Ministry of Education), Fudan University, Shanghai 200433, China

[2]Collaborative Innovation Center of Advanced Microstructures, Nanjing 210093, China

[3]Departamento de Física, Universidade Federal do Ceará, Caixa Postal 6030, Campus do Pici, 60455-900 Fortaleza, Ceará, Brazil

[4]Department of Chemistry, Columbia University, New York, NY 10027, USA

[5]Andlinger Center for Energy and the Environment, Princeton University, New Jersey 08544 USA

[6]Department of Electrical & Computer Engineering, University of Minnesota, Minneapolis, Minnesota 55455, USA

\*Email: hgyan@fudan.edu.cn

## Abstract

Black phosphorus is an infrared layered material. Its bandgap complements other widely studied two-dimensional materials: zero-gap graphene and visible/near-infrared gap transition metal dichalcogenides. Though highly desirable, a comprehensive infrared characterization is still lacking. Here we report a systematic infrared study of mechanically exfoliated few-layer black phosphorus, with thickness ranging from 2 to 15 layers and photon energy spanning from 0.25 to 1.36 eV. Each few-layer black phosphorus exhibits a thickness-dependent unique infrared spectrum with a series of absorption resonances, which reveals the underlying electronic structure evolution and serves as its IR fingerprints. Surprisingly, unexpected absorption features, which are associated with the forbidden optical transitions, have been observed. Furthermore, we unambiguously demonstrate that controllable uniaxial strain can be used as a convenient and effective approach to tune the electronic structure of few-layer black phosphorus. Our study paves the way for black phosphorus applications in infrared photonics and optoelectronics.

**Introduction**

Since the isolation of graphene in 2004[1], tremendous attention has been paid to the family of two-dimensional (2D) materials. Recently, black phosphorus (BP) was reintroduced as a new 2D material[2-5], exhibiting many intriguing properties, such as highly tunable bandgap[6,7], anisotropy[5] and relatively high carrier mobility[2]. It has been predicted that the bandgap of BP is always direct regardless of layer (L) number and ranges from 0.3 eV to 2 eV[6,8], bridging the gap between zero-gap graphene and large-gap transition metal dichalcogenides (TMDs)[9]. Moreover, BP has a puckered hexagonal structure with two nonequivalent directions in the layer plane: armchair (AC) and zigzag (ZZ) (Fig. 1a). Arising from the structural anisotropy, strongly anisotropic mechanical[10], thermal[11], electrical[3,5] and optical[5,6,8,12-15] properties have been highlighted in recent studies, opening up possibilities for conceptually new devices.

Compared to the bulk counterpart, one of the most intriguing distinctions for single or few-layer 2D materials is the highly tunable physical properties, through various techniques. This tunability is typically associated with the modification of the electronic band structure. A nondestructive and accurate characterization technique to monitor few-layer BP band structures is highly desirable, given its predicted strong dependence on layer-thickness[6], stacking order[16], strain[7] and doping[17-19]. Previous photoluminescence (PL)[14] and differential reflectance[20] studies of BP are limited to the visible and near-IR range, and are available only for thin BP layers with layer number less than 5. With the majority of the optical transitions expected in the mid- to near-IR frequency range for few-layer BP, FTIR (Fourier transform infrared spectrometer)-based infrared spectroscopy is believed to be the superior characterization tool. However, up to date, such IR study for mechanically exfoliated few-layer BP (< 15L), with frequency ranging from the mid- to the near-IR, is still lacking.

Here, we systematically investigate the evolution of electronic structures in few-layer BP with layer number ranging from 2 up to 15, and report the experimental demonstration of highly tunable electronic structures in few-layer BP via controllable uniaxial strain[21-25], using polarized IR spectroscopy. For each few-layer BP, the IR spectrum typically exhibits layer-dependent multiple optical resonances and can be readily served as its fingerprints. The IR-absorption shows strong polarization dependence, with strong optical resonances showing up in the AC direction. This dependence provides us a reliable way to determine the crystallographic orientation, which complements polarized Raman spectroscopy. For the latter, however, excitation wavelength and BP thickness complicate the polarization behavior[15,26]. A simple tight-binding model, with only two fitting parameters, can well describe the major optical transitions for all of the measured BP layers. In addition to the main transitions, we also observed unexpected weak absorption features right in the middle of the adjacent main peaks, which we attribute to the forbidden optical transitions. These transitions, expected to be inactive in symmetric BP quantum wells (QW), are made possible by unintentional doping. The physical origin of these weak features has been further confirmed by controlled chemical doping. By collecting the IR spectra for different thickness BP layers, we provide a spectra database, which can be utilized to determine the layer thickness and identify unusual stacking order. More importantly, we also show that few-layer BP is highly sensitive to strain. The bandgap can be modulated by more than 20% with 1% uniaxial strain for a 6L BP. Surprisingly, this effect shows little dependence on the strain direction. Our study demonstrates that IR spectroscopy is an ideal scheme for investigations of band structure engineering through mechanical strain, hydrostatic pressure[27], electrical field[18], magnetic field[28] and chemical doping. The rich band structures of few-layer BP, and their potential large tunability, promise a wide range of applications in IR photonics and optoelectronics[29], such as polarization-sensitive photo-detectors[12,30], modulators, strain sensors and IR lasers.

## Results

**Sample preparation and polarization resolved IR spectroscopy.** Few-layer BP flakes were prepared by mechanical exfoliation of bulk crystals (HQ Graphene, Inc), with areas varying from several hundred of to ten thousand $\mu m^2$. To minimize the influence of Fabry-Perot interference of the substrate, thick quartz substrates (~0.3 mm in thickness) were used to support BP flakes. Sample thickness was first determined from the optical contrast[31] under a microscope, and then was further verified by IR absorption spectra, as we will discuss later. According to our optical contrast analysis of many BP flakes (more than 100 few-layer flakes) with different thickness, we were able to quite accurately determine the thickness of BP flakes, especially for thinner ones (see Supplementary Fig. 1). To minimize the effect of sample degradation in air, IR measurements were typically finished within 1 hour after sample exfoliation. The IR transmission (extinction) spectra with photon energy from 0.25 eV to 1.36 eV were obtained using a FTIR in conjunction with an IR microscope at room temperature (see Methods). The lower bound cutoff photon energy is restricted by the quartz substrate. Fig. 1b is an optical image of a representative 6L sample, with areas of ~5000 $\mu m^2$, large enough for us to obtain accurate IR extinction spectrum. For an atomically thin layer material sitting on a thick transparent substrate, the extinction (1-$T/T_0$) is directly proportional to the real part of the optical conductivity $\sigma(\hbar\omega)$ [32,33], where $T$ and $T_0$ donate the transmittance of samples on substrate and bare substrates, respectively.

We measured polarization-resolved IR absorption spectra of the 6L sample, with normal light incidence and polarization angles ranging from 0° to 360° in steps of 15° (Supplementary Fig. 2). For clarity, only six spectra are shown in Fig. 1d. For such large area samples, the measured extinction can be directly translated into the real part of the optical conductivity according to the formula[32,33] $\text{Re}\,\sigma(\hbar\omega) = (1 - T/T_0) \cdot (n_s + 1) \cdot c/8\pi$, where $n_s$ is the refractive index of the quartz substrate and $c$ is the speed of light. Two prominent peaks (labeled as $E_{11}$ and $E_{22}$) can be identified, revealing rich features of the underlying electronic structures. Due to the confinement along the $z$ direction, conduction bands and valence bands split into multiple subbands with well-defined quantum numbers $n$, and optical transitions are allowed only from valence subbands to conduction subbands with the same index $n$ for normal light incidence[17,18]. In low dimensional materials, it's well known that the optical resonance peaks arise from excitons, due to significantly reduced screening of Coulomb interactions[34-36]. However, for clarity, we adopt the band to band transition picture, rather than the excitonic picture when we refer to the measured IR resonance peaks. We assign the two peaks $E_{11}$ and $E_{22}$ to optical transitions $v_1 \rightarrow c_1$ and $v_2 \rightarrow c_2$, respectively, illustrated in the left panel of Fig. 1c. Higher order subband transitions are beyond the limit of our measurement range. Previous theory[6,8,17] and experiments[5,12] have demonstrated strong anisotropy in IR absorption for thick BP layers: BP absorbs light polarized along the AC direction, with much less absorption for light with polarization along the ZZ direction, because mirror symmetry in the $x$-$z$ plane forbids optical transitions for the ZZ polarization. Therefore, the optical conductivity of BP maximizes in the AC direction, as seen in Fig. 1e. By rotating the linear polarizer, the AC direction is determined to be ~2° relative to the pre-selected $x$ direction. This unique anisotropy makes IR spectroscopy a very convenient and unambiguous method to determine the crystallographic orientation. For the peak $E_{22}$, Re($\sigma$) reaches a maximum of 2.6 $\sigma_0$ ($\sigma_0 = \pi e^2 / 2h$ is the renowned universal optical conductivity of graphene[33,37]). This conductivity gives an absorption of ~6 % at $E_{22}$ for a suspended 6L BP, indicating a 1 % absorption per BP layer. The most interesting feature is the weak peak between $E_{11}$ and $E_{22}$, labeled with an asterisk (*). We assign it to the transitions between $v_1$ and $c_2$, and $v_2$ and $c_1$, with transition energies denoted as $E_{12}$ and $E_{21}$ respectively, as shown in the right panel of Fig. 1c. Because these transitions involve bands with different quantum numbers[18,38], we refer them as hybrid transitions. More detailed discussions will be given later.

**The evolution of the electronic structures with layer number.** We performed polarization-resolved IR absorption measurements on few-layer BP from 2L to 15L, as shown in Figs. 2a-j. For comparison, we also measured the absorption spectrum for a bulk BP (Fig. 2k, thickness > 100 nm). All the experiments were performed at room temperature under ambient conditions. The optical features of monolayer BP lie in the visible frequency range, beyond the IR measurement limit[3,14]. As seen in Supplementary Fig. 3, the bandgap PL emission peak is around 1.67 eV. Except for the 6L and a couple of other samples, the sample size may be smaller than the IR beam size. As a result, the light extinction values are usually smaller than the true value and comparison of the absolute extinction values for different layers in Fig. 2 is not meaningful. Nevertheless, the peak positions are very informative. For the 2L sample, a salient peak around 1.14 eV is observed with incident light polarized along the AC direction (black line in Fig. 2a). We assign this peak to the lowest energy transition $v_1 \rightarrow c_1$, labeled as $E_{11}$. As predicted by theory[17], the spectrum in the ZZ direction is featureless (red line in Fig. 2a). With the increase of layer number, the peak $E_{11}$ undergoes a monotonous red shift from 1.14 eV (2L) to 0.38 eV (15L), and eventually reaches 0.34 eV in the bulk limit. The observed behavior qualitatively agrees with previous theoretical predictions, in which strong interlayer interaction is responsible for the bandgap decrease[6,8,39].

Interestingly, a new series of peaks appear in the absorption spectra for few-layer BP, in addition to the lowest energy peaks. All of the higher energy peaks exhibit similar polarization dependence to that of $E_{11}$ transitions. For the 5L sample, the transition $E_{22}$ ($v_2 \rightarrow c_2$) peak is at 1.30 eV. Up to 15L, the peak $E_{22}$ red-shifts to 0.48 eV, showing a similar layer dependence to that of the peak $E_{11}$. For the 9L sample, the third peak (labeled as $E_{33}$) starts to emerge at around 1.22 eV, assigned to the third transition $v_3 \rightarrow c_3$. Again, $E_{33}$ shifts toward lower energies as the layer number increases. For the 13L and 15L samples, at least four peaks can be identified unambiguously. Apparently, each few-layer BP has a unique IR spectrum and is readily distinguishable from each other, even for relatively thick samples such as 15L ones. For thinner BP flakes (typical layer number is less than five), sample thickness can be accurately determined by the $E_{11}$ peak position. Although the $E_{11}$ peak shows very little difference for thicker samples, the higher energy transition peaks ($E_{22}$, $E_{33}$, $E_{44}$) show more discernable differences. Therefore, it's quite accurate to determine the BP thickness by IR spectroscopy with atomic-level precision, given the clear sequence of subband transitions.

For *N*-layer BP, the conduction and valence bands both split into *N* 2D subbands due to the layer-layer interactions[40]. These additional peaks originate from optical transitions between higher index subbands, revealing a QW-like band structure[17]. Following a tight binding model to take into account the nearest-neighbor layer-layer interactions (see Supplementary Note 1)[40], at Γ point of the 2D Brillioun zone, the transition energy between valence subband and conduction subband with the same index *n* (*n* = 1,2,3,...,*N*) is given by

$$E_{nn}^{N} = E_{g0} - 2(\gamma_c - \gamma_v)\cos(\frac{n\pi}{N+1}) \qquad (1)$$

where $E_{g0}$ is the bandgap of monolayer BP, $\gamma_c$ and $\gamma_v$ characterize the nearest interlayer couplings for the conduction and valence bands, respectively. Equation (1) describes the series of resonance peaks in IR spectra. As seen in Fig. 3a, all of the four branches ($E_{11}$, $E_{22}$, $E_{33}$, $E_{44}$) of optical transitions can be well fitted simultaneously, with fitting parameters $E_{g0}$ = 2.12 eV, $\gamma_c$ - $\gamma_v$ = 0.88 eV. For completeness, we also put a PL data point (hollow square) for a monolayer BP in the figure. With increasing layer number *N*, the energy spacing between two adjacent subbands within the conduction or valence bands decreases monotonically. In the bulk limit, these subbands are so closely spaced that they evolve into quasi-continuous bands at last. Therefore, the spectrum for bulk BP in Fig. 2k has no resolvable peaks. Our results reveal the crossover of band structure from 2D to 3D. For the 13L and 15L BP, the relationship between transition energy $E_{nn}$ and subband index $n$ (or quantum number) is well described by the QW based formula[17] $E_{nn} = a + b \cdot n^2$, where $a$ and $b$ are fitting parameters, as shown in Fig. 3b. This is a direct consequence of the QW confinement and can also be derived from equation (1), as shown in the Supplementary Note 1.

The spectra set in Fig. 2 is a valuable fingerprint database for few-layer BP. Spectra with dramatic deviation from those typically signal large structural change. For instance, as predicated by theory, the stacking order can affect the band structure of BP layers[16]. Indeed, we observed a few anomalous spectra and attributed them to BP layers with different stacking order (See Supplementary Fig. 4).

In single and few-layer BP, it has been shown theoretically that the optical transitions may arise from excitons[6,41]. In the meantime, a theoretical calculation without taking into account excitonic effect also gives insightful results[17]. In our description of the optical resonance energy, we don't take into account electron-electron and electron-hole (excitonic effect) interactions and adopt a simple tight binding model. Interestingly, the model is very consistent with our results. However, this doesn't preclude the possible many-body effects in few-layer BP. Previously, for other low-dimensional materials, such as 1D carbon nanotubes [34], 2D TMDs[36,42] and graphene[43], it was reported that single particle band structure renormalization due to electron-electron interactions, which typically increases the optical transition energies, are largely cancelled out by the exciton binding energy. As a result, the experimentally observed optical transition energy can often be accounted for within the single particle description. Very likely, this is also the case for few-layer BP.

**Strain engineering of the electronic structures.** The electronic structure of mono- and few-layer BP has been predicted to be very sensitive to strain[7,21-24,44]. Recently, optical spectroscopy in the visible frequency range showed strong spectral variation for wrinkled multilayer BP, indicating an inhomogeneous strain field[45]. Raman spectroscopy has been utilized to study the lattice vibrations of few-layer BP under uniaxial strain and strong anisotropy has been demonstrated[46,47]. In our experiment, we transferred few-layer BP on flexible PET (polyethylene terephthalate) substrate and employed a two-point bending apparatus to induce controllable uniaxial strain ($\varepsilon$), as illustrated in Fig. 4a. We monitored the electronic structure change of the strained few-layer BP by IR spectroscopy. The strain setup is detailed in Supplementary Note 2. Before applying strain, the crystallographic orientation of the BP flake was identified by polarized IR spectroscopy, so that the uniaxial strain can be applied to the sample in the desired directions (AC and ZZ directions). To avoid the complication from sample slippage, the applied strain was typically kept below 1% during the entire strain process. Within such moderate strain range, the stretch process is reversible and repeatable. Fig. 4c shows the IR extinction spectra of a representative 6L BP sample under varying tensile strains up to $\varepsilon$ = 0.92%, with strain applied along the AC and ZZ directions respectively. For clarity, the spectra are vertically shifted. As the tensile strain increases gradually, the two characteristic peaks of $E_{11}$ and $E_{22}$ both blueshift monotonically.

As we know, BP is a highly anisotropic material, especially its mechanical properties exhibit strong crystallographic orientation dependence. For example, monolayer BP has been predicted to contract in the out of plane direction under tensile strain along the AC direction (positive Poisson's ratio), while it expands under ZZ tensile strain (negative Poisson's ratio)[48]. In addition, the Young's modulus of monolayer BP in the ZZ direction is predicted to be 3.8 times larger than that in the AC direction[10]. With these in mind, let's now examine the direction dependence of the uniaxial strain effect on the electronic structures. The peak positions in Fig. 4c are summarized in Fig. 4d as a function of tensile strain in both AC and ZZ directions. Surprisingly, little difference is observed for the two distinct strain directions. For each strain, the peak positions of the sample stretched in both directions are almost always the same. From the linearly fitted lines, we can extract that the blueshift rate for $E_{11}$ is 117 meV/% (AC direction) and 124 meV/% (ZZ direction), respectively, and 99 meV/% for $E_{22}$ in both directions. It's not a coincidence for this specific sample. We have performed similar measurements on multiple samples, no strain direction dependence beyond experimental uncertainty has been observed, in sharp contrast to the giant anisotropic Raman response to uniaxial strain[46,47].

Within the strain range from $\varepsilon$ = 0% to 0.92%, the optical bandgap ($E_{11}$) can be continuously tuned from 0.54 eV to 0.65 eV for this 6L sample. Such large tunability is highly desirable for the application in high-sensitivity strain sensors. From the extracted peak shifting rates, we conclude that 1% tensile strain leads to a 23% increase of the bandgap for a 6L BP. The fractional change of the bandgap is even greater for thicker BP samples, given that the bandgap is smaller and the shift rate is similar for all layer thickness, as discussed later on. These observations are consistent with previous first principle calculations for monolayer and bulk BP[21-24].

The observed electronic structure evolution can be understood within the tight-binding framework[25,39,44,49]. Since strain is applied in the basal plane, it mostly affects the in-plane bonding. As a consequence, the strain effect (shift rates of the optical transitions) will be similar for mono- and few-layer BP, with little dependence on the layer thickness. Indeed, we observed a similar shift rate as that of the 6L BP sample for the $E_{11}$ transition of a 3L BP sample, as shown in Supplementary Fig. 7. In fact, such phenomenon can be inferred from equation (1). In-plane strain has strong effect on $E_{g0}$ and smaller effect on $\gamma_c$ - $\gamma_v$, which renders all optical transitions $E_{nn}^N$ similar strain dependence. Therefore, we only need to consider the strain effect on monolayer BP, whose band structure around the band-edge can be quite satisfactorily captured by two hopping parameters $t_1^{//}$ and $t_2^{//}$ [39,50], as schematically illustrated in Fig. 4b. The values of $t_1^{//}$ and $t_2^{//}$ have been obtained theoretically through DFT calculations and tight-binding parameterization[39]. The bandgap of monolayer BP is expressed as $E_{g0} \approx 4t_1^{//} + 2t_2^{//}$, with $t_1^{//}$ < 0 and $t_2^{//}$ > 0. With this expression, we can immediately explain why it increases with tensile strain along the ZZ direction. When a small tensile strain is applied along the ZZ direction, the amplitude of the hopping parameter $|t_1^{//}|$ will decrease because the relevant bond length increases. At the same time, $|t_2^{//}|$ has no change due to the fact that the relevant bond length remains unchanged, given that the bond orientation is perpendicular to the strain direction. Consequently, the bandgap $E_{g0}$ increases, given that $t_1^{//}$ is negative in the first place. A qualitative argument cannot be applied for strain along the AC direction, because both $|t_1^{//}|$ and $|t_2^{//}|$ decrease in this case. As demonstrated later, the overall amplitude of the bandgap still increases due to the dominant change of $|t_1^{//}|$ over $|t_2^{//}|$ in $E_{g0}$. The increase of the bandgap under uniaxial tensile strain for BP layers is in sharp contrast to mono- and few-layer $MoS_2$, whose bandgap shrinks[51].

More interestingly, the strain along the ZZ and AC directions have quantitatively the same effect on the optical transitions. This is very counter-intuitive, given the fact that BP is so anisotropic in terms of almost every property, particularly for the mechanical properties. In order to better understand this behavior, we have to examine the strain effect more quantitatively within the tight-binding framework. With the hopping parameters obtained from DFT calculations[39] (at zero strain, $t_1^{//}$ = -1.22 eV , $t_2^{//}$ = 3.665 eV ), and, as a common practice, assume that they are inversely proportional to $r^2$ ($r$ is the bond length), one can directly derive the bandgap dependence on the strain for monolayer BP[44,49]:

$$\Delta E_{g0} = 4.1\varepsilon_x + 5.7\varepsilon_y - 12.9\varepsilon_z \qquad (2)$$

with the unit of eV. In the equation, $\varepsilon_x$, $\varepsilon_y$, $\varepsilon_z$ is the strain along *x*, *y* and *z* direction, respectively. It should be noted that for monolayer BP, $\varepsilon_z$ reflects the change of the thickness of the single puckered sheet, i.e., the vertical distance between atoms on the top and atoms at the bottom of the puckered sheet. In fact, in another 2D material - monolayer graphene, $\varepsilon_z$ has no meaning since all atoms are on the same plane. According to equation (2), it's obvious that the pure ZZ strain ($\varepsilon_y$) is more effective to tune the bandgap than the pure AC strain ($\varepsilon_x$), which is consistent with our expectation. However, experimentally, due to the Poisson effect of the PET substrate, the strain is not truly uniaxial and deformations in other directions exist as well. BP layers stick to the underlying substrate and they deform in the same way as the substrate in the *x-y* plane. More specifically, the BP layer stretched in ZZ direction will slightly shrink in AC direction and vice versa. However, because the Poisson effect of the substrate has no orientation dependence, this effect cannot smear out the strain direction dependence shown in equation (2) and stretch along ZZ direction is still more efficient to change the electronic structure. On the other hand, in $z$ direction, BP layers are free to shrink or expand and $\varepsilon_z$ is nonzero. We attribute the observed lacking of orientation dependence for the strain effect to the change of the monolayer thickness under in-plane strain. More specifically, for a monolayer BP, the tensile strain along the AC direction leads to compression in $z$ direction, which gives rise to additional bandgap enhancement (see equation (2), noticing that the coefficient of $\varepsilon_z$ is negative). On the contrary, tensile strain along the ZZ direction results in expansion in the $z$ direction due to a predicted negative Poisson's ratio ($\nu_{zy} < 0$)[48], which partially cancels out the zigzag strain effect. More quantitatively, through equation (2), we find out that a Poisson's ratio difference ($\nu_{zx} - \nu_{zy}$) of ~0.12 in $z$ direction will smear out the difference of the effect for strain along the ZZ and AC directions. This is consistent with DFT calculations, which have obtained $\nu_{zx}$ and $\nu_{zy}$, with their difference ranging from 0.07 to 0.3[24,48]. It should be noted that the different Poisson's ratios in $z$ direction for strain along the ZZ and AC directions are direct consequences of the hinge-like structure of a BP layer[48].

Now we see that equation (2) semi-quantitatively describes the strain dependence of the electronic transitions in mono- and few-layer BP. However, the derived shift rate is smaller than what we observed experimentally. This discrepancy requires more effort in exactly determining the hopping parameters $t_1^{//}$ and $t_2^{//}$ and reassessment of the assumption that $t \propto 1/r^2$. If we keep $t_1^{//}$ and $t_2^{//}$ values provided by Rudenko et al.[39] and assume that $t \propto 1/r^\beta$, we find out that the index $\beta \sim 6$ can quantitatively account for the measured shift rates[25].

## Discussion

In addition to the main peaks in Fig. 2, there are weak absorption peaks between $E_{11}$ and $E_{22}$ transitions, indicated by asterisks (*). For 4,5,6-layer BP, the weak feature shows one bump, while for 7,8,9-layer case, it splits into two peaks, as indicated by double asterisks. We assign these weak peaks to the hybrid transitions from the valence subband to the conduction subband with different quantum numbers: $v_1$ to $c_2$ ($E_{12}$ transition) and $v_2$ to $c_1$ ($E_{21}$ transition), as illustrated in the right panel of Fig. 1c. From the energy diagram in Fig. 1c, it's straightforward to have $(E_{11} + E_{22})/2 = (E_{12} + E_{21})/2$, which means that the average peak position of the $E_{12}$ and $E_{21}$ transitions is right in the middle of the $E_{11}$ and $E_{22}$ transitions. Indeed, this is exactly what we observed. Fig. 5a plots the average peak position of $E_{12}$ and $E_{21}$ as a function of $(E_{11} + E_{22})/2$. For the not well-split peaks below 6L, $(E_{12} + E_{21})/2$ is just the peak position of the single bump. In addition to the data points extracted from the spectra in Fig. 2, other data points for additional 7L or 8L samples are also shown. The solid line is a straight line with $(E_{11} + E_{22})/2 = (E_{12} + E_{21})/2$. We see that data points all follow the line, which is consistent with hybrid transitions.

The origin of these hybrid transitions deserve some discussion. Ideally, these hybrid transitions are forbidden and the optical oscillator strength should be zero for a symmetric QW with a normal-incident IR beam[38]. We believe that the unintentional doping from the substrate and/or air can break the symmetry of the BP QWs and relax the selection rules[38,52]. This is consistent with a theoretical study, which shows doping indeed can activate such transitions[18]. To verify this conjecture, we intentionally doped a 9L sample with nitric acid vapor. Such doping scheme has been proved very efficient to increase hole carrier density in graphene[53]. Fig. 5b shows the extinction spectra before and after doping. Clearly, the hybrid transition intensity becomes almost 2 times as strong after treatment, as shown in Fig. 5c. Meanwhile, $E_{11}$ slightly redshifts and its oscillator strength decreases dramatically, which can be understood as due to reduced wavefunction overlap for valence and conduction states[18]. These behaviors are all consistent with doping effect[18]. This controlled doping experiment suggests that doping can relax the optical transition selection rules and make forbidden transitions possible. It's natural to ask whether sample degradation makes hybrid transitions observable, since BP flakes are sensitive to oxygen and water in air and degrade at ambient condition[54]. In order to examine the degradation effect, we monitored the spectra for samples as a function of time. We find that the degradation is a slow process and freshly cleaved few-layer samples (L > 3) can keep the same spectra for several hours. However, thinner samples (L < 3) degrade much faster and spectra can change in two hours after cleavage. The spectra shown in Fig. 2 were typically measured within one hour after cleavage and we believe the degradation effect on the spectra is minimal. Supplementary Fig. 5 presents two spectra for the 4L sample right after cleavage and after one day. A blueshift of the main peak $E_{11}$ can be observed, presumably due to defects introduced by the degradation[55]. In addition, the hybrid peak (marked with *) disappears, suggesting that degradation is not the origin of the observed hybrid transitions. It should be noted that for some cases, even freshly cleaved BP samples don't show the hybrid peaks, presumably due to a low doping concentration.

Though with small oscillator strength, the hybrid transitions are very informative. With both the main peaks and hybrid peaks showing up, the energy levels of $v_1$, $v_2$, $c_1$, $c_2$ (labeled as $E_{v1}$, $E_{v2}$, $E_{c1}$, $E_{c2}$) with respect to a common reference energy can in principle all be determined. For 4,5,6-L BPs, the splitting between $E_{12}$ and $E_{21}$ are tiny and they are nearly degenerate, which indicates that the energy spacing between $v_1$ and $v_2$ ($E_{v1} - E_{v2}$), $c_1$ and $c_2$ ($E_{c2} - E_{c1}$) are almost identical and electron and hole are nearly symmetric in $z$-direction[17]. For those BP layers with split hybrid transitions (L > 6), the energy spacings $E_{v1} - E_{v2}$ and $E_{c2} - E_{c1}$ are different, which indicates an electron-hole asymmetry in the $z$-direction. Previous cyclotron resonance experiments for bulk BP showed that the effective mass in $z$-direction for electrons and holes are very different, with hole mass almost 2 times of electron mass[28]. This is consistent with our observations that the BP layers with larger thickness have split hybrid peaks and shows more pronounced electron-hole asymmetry. The reason why thin BP layers (L < 6) have almost degenerate $E_{12}$ and $E_{21}$ transitions merits further theoretical and experimental investigations. Meanwhile, the linewidths and relative intensities of $E_{12}$ and $E_{21}$ transitions, as well as their doping dependence deserve careful examinations as well. These studies will give valuable information concerning the band structure, especially electron-hole asymmetry in few-layer BP samples.

## Methods

**Sample preparation.** Few-layer BP samples were prepared by a PDMS (polydimethylsiloxane) assisted mechanical exfoliation method. In brief, BP flakes were first cleaved on a PDMS substrate, then transferred onto a quartz substrate once thin flakes were identified under microscope. After the transfer, sample thickness was determined using a Nikon inverted microscope (Eclipse Ti-U), in combination with IR spectroscopy, as mentioned in the main text.

**Polarized IR spectroscopy.** Polarization-resolved IR spectroscopy was performed using a Bruker FTIR spectrometer (Vertex 70v) integrated with a Hyperion 2000 microscope. A combination of tungsten halogen lamp and globar was used as light sources to cover the wide energy range from mid- to near-IR (0.1–1.36 eV). The incident light was focused on BP samples with a 15X IR objective, the polarization was controlled by a broadband ZnSe grid polarizer. IR radiation was collected by a liquid nitrogen cooled MCT detector. Generally, aperture size was set to be larger than the sample size, in order to improve signal to noise. All the measurements were conducted at room temperature in ambient conditions.

**Chemical doping of few-layer BP samples.** The intentional doping for the 9L sample shown in Fig. 4 was done through $HNO_3$ vapor exposure. The sample was exposed for 5 seconds. Typically, longer time of exposure will gradually damage the sample.

**Data availability.** The data that support the findings of this study are available from the corresponding author upon request.

**Acknowledgements**

H. Y. is grateful to the financial support from the National Young 1000 Talents Plan, the National Key Research and Development Program of China (Grant number: 2016YFA0203900), and Oriental Scholar Program from Shanghai Municipal Education Commission. We thank Profs. Yuanbo Zhang, Zhengzong

Sun, Lei Shi and Jun Shao for experimental assistance and Profs. Feng Wang, Xingao Gong, D. R. Reichman, F. M. Peeters for stimulating discussions. A.C. Acknowledges financial support by CNPq through the PRONEX/FUNCAP, PQ and Science Without Borders Programs, and the Lemann Foundation. Part of the experimental work was carried out in Fudan Nanofabrication Lab.

**Author contributions**

H. Y. and G. Z. initiated the project and conceived the experiments. G. Z., S. H., C. S. prepared the samples. G. Z. performed the measurements and data analysis with assistance from S. H and C. S.. T.L., A. C. and V. O. provided modeling and theoretical support. H.Y. and G. Z. co-wrote the manuscript with inputs from T. L.. H. Y. supervised the whole project. All authors commented on the manuscript.

**Competing financial interests:** The authors declare no competing financial interests.

**Additional information**

Supplementary information accompanies this paper at www.nature.com. Reprints and permission information is available online at http://npg.nature.com/reprintsandpermissions/. Correspondence and requests for materials should be addressed to H.Y. (hgyan@fudan.edu.cn).

**Figure captions**

**Figure 1 | Few-layer BP samples and polarization dependent IR spectroscopy.** (**a**) Lattice structure of few-layer BP, showing the puckered hexagonal crystal with two nonequivalent directions: armchair (AC, *x*-direction) and zigzag (ZZ, *y*-direction). (**b**) Optical image of a representative 6L sample. Scale bar: 20 μm. (**c**) Schematic illustrations of optical transitions between quantized subbands, solid and dotted arrows indicate the main transitions ($\Delta n = 0$) and hybrid transitions ($\Delta n = \pm 1$), respectively. (**d**) Real part of the optical conductivity $\sigma$ for the 6L sample in Fig. 1b, in the unit of the universal optical conductivity $\sigma_0 = \pi e^2/2h$, with polarization angles of incident light from 15° to 90°, in steps of 15°. The marks $E_{11}$ and $E_{22}$ donate the first and second subband transitions, asterisks (*) donate hybrid transitions. (**e**) Real part of the optical conductivity $\sigma$ at peaks $E_{11}$ and $E_{22}$, as a function of the polarization angle $\theta$. The solid lines are $\cos^2\theta$ fits.

**Figure 2 | Layer dependent IR spectrum.** (**a**)-(**j**) Extinction spectra ($1-T/T_0$) for few-layer BP on quartz substrates with layer number $N = 2$-9, 13 and 15. The labels $E_{11}$, $E_{22}$, $E_{33}$ and $E_{44}$ donate the first, second, third and fourth subband transitions, asterisks (*) donate hybrid transitions, respectively. (**k**) Extinction spectra for a thick bulk BP (thickness > 100 nm) on intrinsic Si substrate. The black and red curves represent spectra for two different light

polarizations.

**Figure 3 | Evolution of the electronic structure in few-layer BP.** (**a**) Peak energies of the first, second, third and fourth transitions between quantized subbands as a function of layer number $N$. The solid lines represent theory predictions according to the quasi-1D tight binding model. The black hollow square is extracted from the PL spectroscopy of monolayer BP. The grey horizontal line in the upper panel is the upper limit of our IR measurement range, while the one in the lower panel is the measured bandgap of bulk BP. (**b**) Peak energies as a function of quantum number $n$ for the 13L and 15L BP samples, fitted by $E_{nn} = a + b*n^2$, where $a$ and $b$ are fitting parameters.

**Figure 4 | Strain engineering of the electronic structures.** (**a**) Schematic illustration of the two-point bending apparatus using a flexible PET substrate. (**b**) Schematic illustration of two in-layer hopping parameters ($t_1^{//}$ and $t_2^{//}$) and one out-of-plane hopping parameter ($t^{\perp}$) in a bilayer BP. (**c**) Extinction spectra (1-$T/T_0$) of a 6L BP sample under varying tensile strains, with strain applied along the armchair (red) and zigzag (blue) directions. The spectra are vertically offset. 0.92%$x$ (0.92%$y$) indicates applying 0.92% strain along the armchair (zigzag) direction. The incident light is polarized along the armchair direction. The dashed lines trace the shift in the transition energies. (**d**) The $E_{11}$ and $E_{22}$ peak energies as a function of tensile strains, the strain direction is along the

armchair (red) and zigzag (blue) directions, respectively. The solid lines are linear fits to the data.

**Figure 5 | Hybrid transitions.** (**a**) Comparison between the average value of main peaks ($E_{11} + E_{22}$)/2 and that of hybrid peaks ($E_{12} + E_{21}$)/2 for BP layers with different thickness, the red line $y = x$ serves as a guide to the eye. (**b**) Extinction spectrum of a 9L BP before (black) and after (red) chemical doping through $HNO_3$ vapor treatment. (**c**) Enlarged view of the hybrid peaks between $E_{11}$ and $E_{22}$ in Fig. 5b, with tilted backgrounds removed for clarity.

# Figure 1: Few-layer BP samples and polarization dependent IR spectroscopy

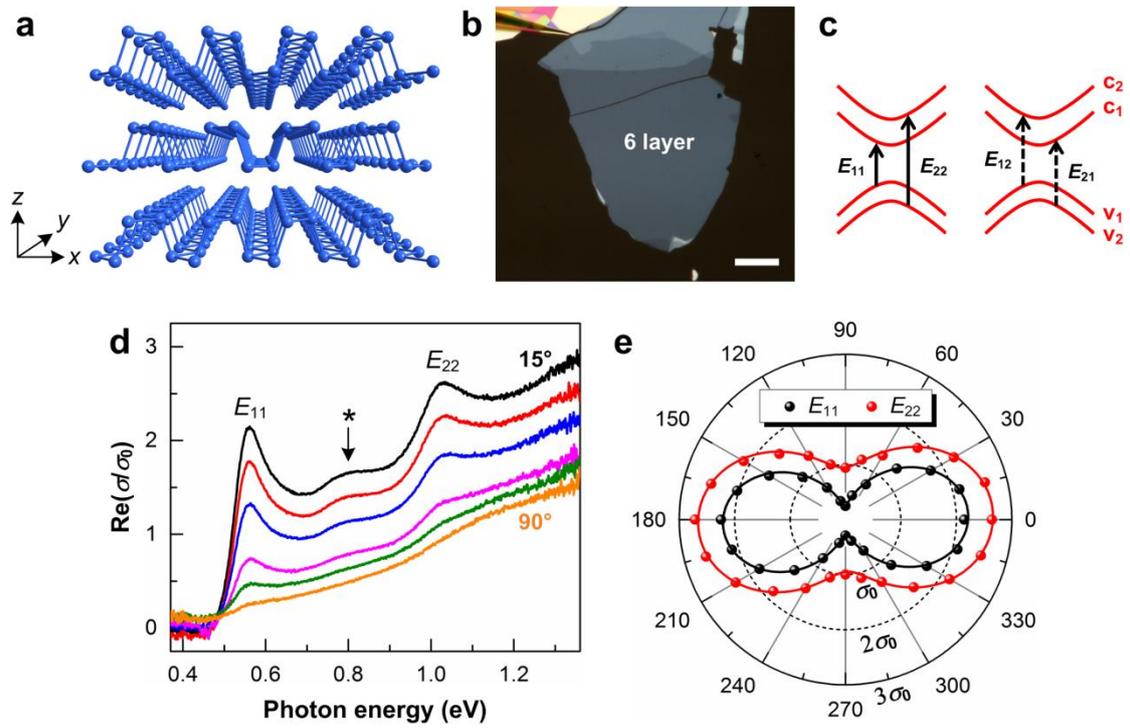

# Figure 2: Layer dependent IR spectrum

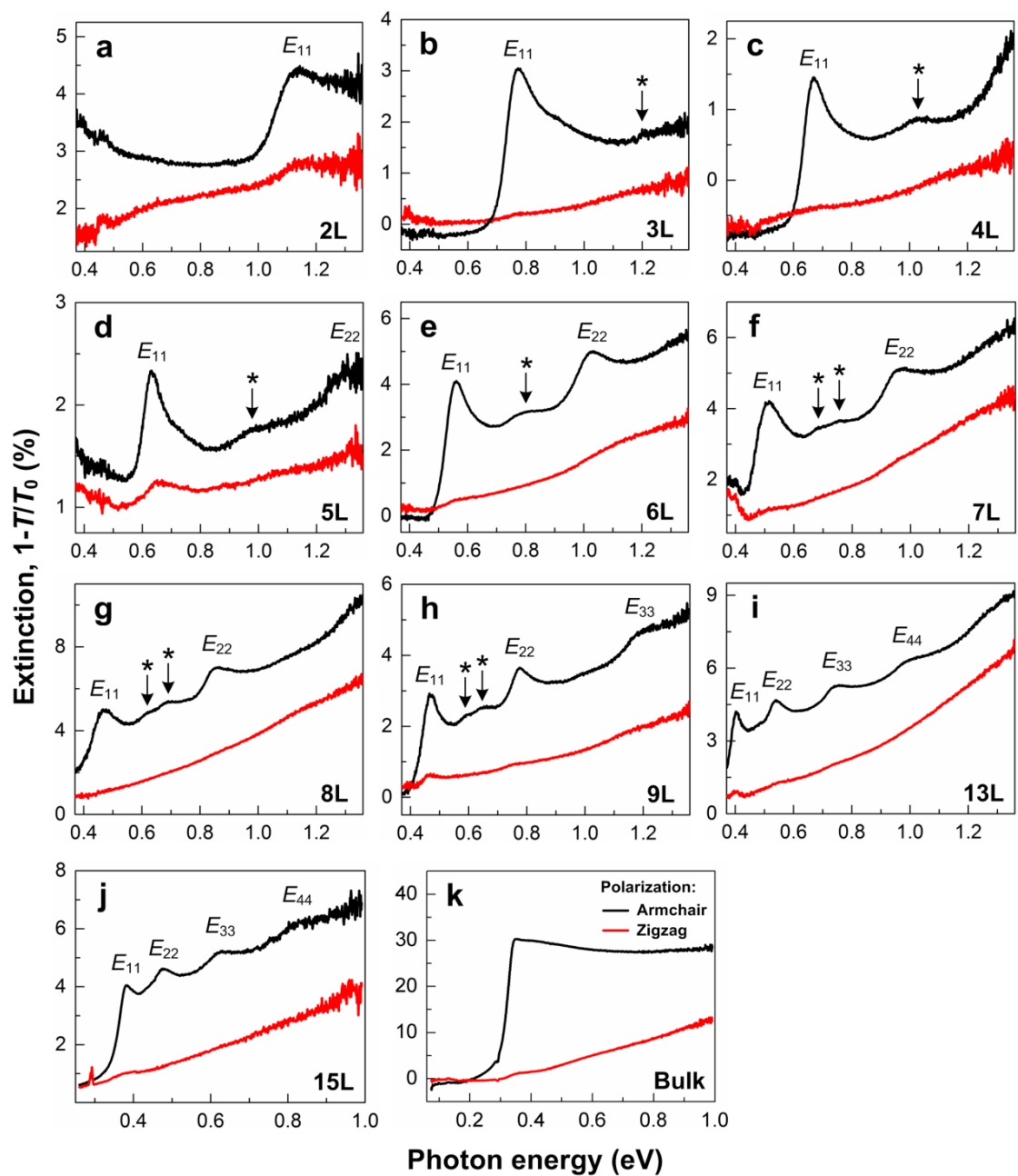

**Figure 3: Evolution of the electronic structure in few-layer BP**

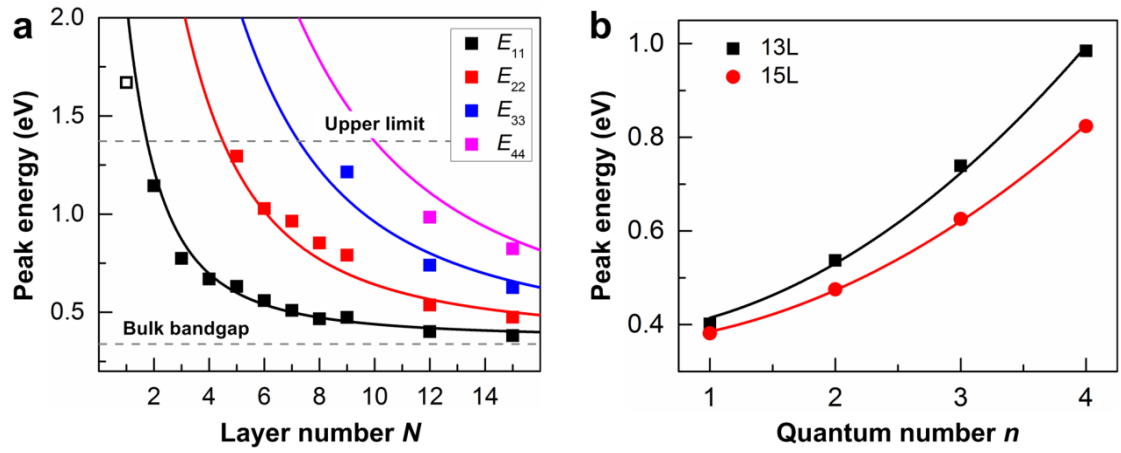

# Figure 4: Strain engineering of the electronic structures

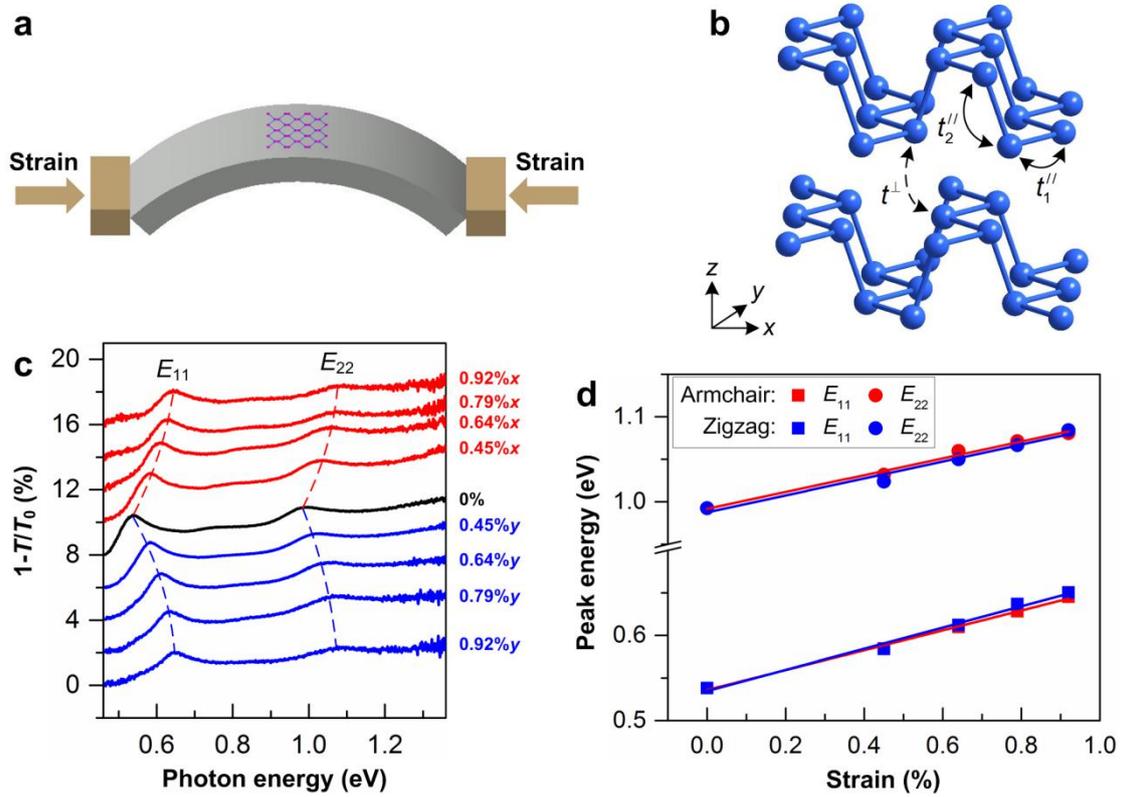

# Figure 5: Hybrid transitions

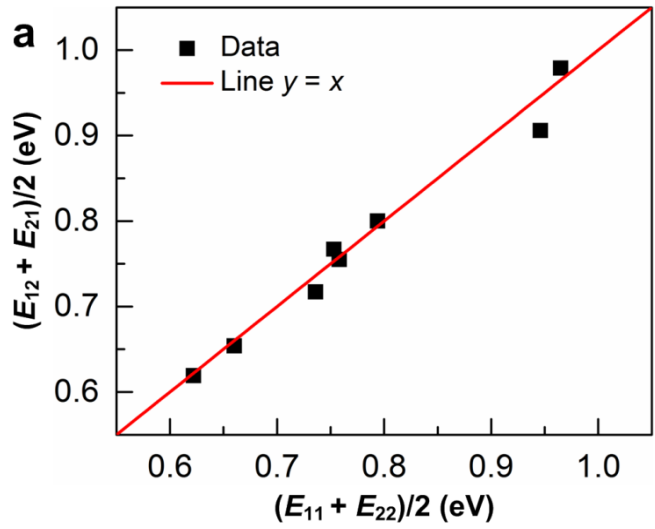
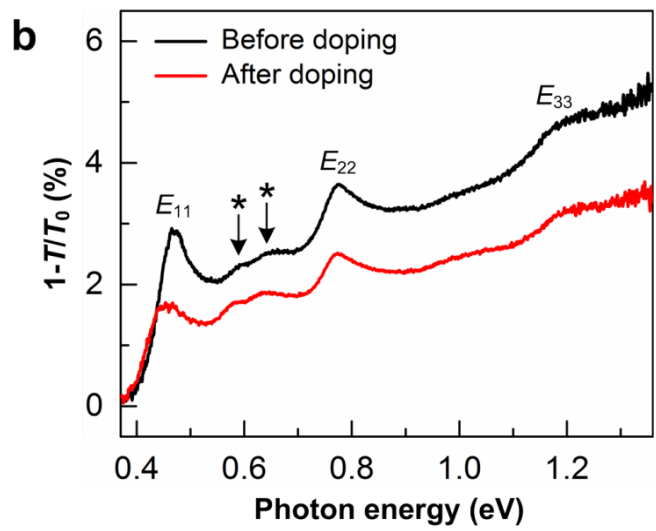
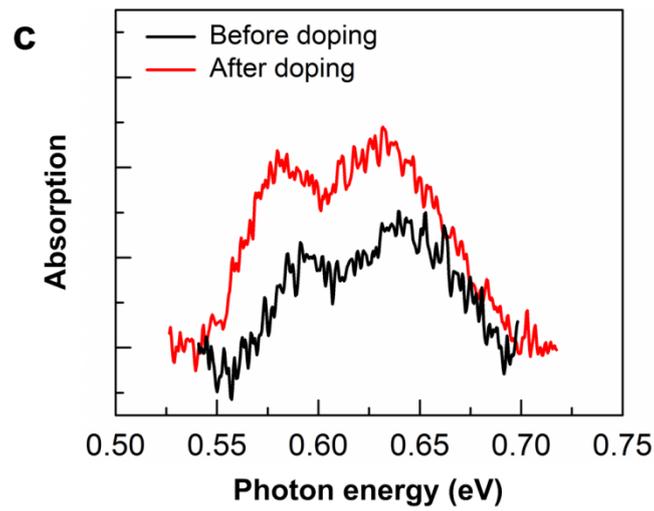